\documentclass[twocolumn,showpacs,preprintnumbers,amsmath,amssymb]{revtex4}

\usepackage{graphicx}
\usepackage{dcolumn}
\usepackage{subfigure}
\usepackage{epsfig}
\usepackage{pstricks,pst-grad}
\usepackage{fancyvrb}
\usepackage{mathptmx}
\usepackage{pslatex}

\def\ket #1{\vert #1\rangle}

\newcommand{\dm}[2]{\ensuremath{|#1\rangle\langle #2|}}
\newcommand{\four}[6]{\ensuremath{\frac{\text{c}_{#1}{#2}\text{c}_{#3}}{\text{c}_{#4}{#5}\text{c}_{#6}}}}

\newcommand{\e}{\mathrm{e}}
\renewcommand{\i}{\mathrm{i}}
\newcommand{\SO}{\mathrm{SO}}
\newcommand{\SU}{\mathrm{SU}}

\begin{document}

\title{Tight Noise Thresholds for Quantum Computation with Perfect Stabilizer Operations}

\author{Wim van Dam}
\email{vandam@cs.ucsb.edu}
\thanks{}
\affiliation{%
Department of Computer Science, University of California, Santa Barbara, CA 93106, USA\\
Department of Physics, University of California, Santa Barbara, CA 93106, USA
}
\author{Mark Howard}
\email{mhoward@physics.ucsb.edu}
\affiliation{%
Department of Physics, University of California, Santa Barbara, CA 93106, USA
}%
\date{\today}

\begin{abstract}
We study how much noise can be tolerated by a universal gate set before it  loses its quantum-computational power. Specifically we look at circuits with perfect stabilizer operations in addition to imperfect non-stabilizer gates.
We prove that for all unitary single-qubit gates there exists a tight depolarizing noise threshold that determines whether the gate enables universal quantum computation or if the gate can be simulated by a mixture of Clifford gates. This exact threshold is determined by the Clifford polytope spanned by the $24$ single-qubit Clifford gates.
The result is in contrast to the situation wherein non-stabilizer qubit states are used; the thresholds in that case are not currently known to be tight.
\end{abstract}

\pacs{03.67.-a, 03.67.Pp}
\maketitle

\paragraph*{Introduction.}

A way to study the resources needed for universal quantum computation (\textsc{uqc}) is to analyze the transition from a system that can provide \textsc{uqc} to one that is classically efficiently simulable.  A particularly useful example of such a system is given by the stabilizer operations, which are made by a combination of: preparation of $\ket{0}$ states, unitary Clifford gates, measurements in the $\{\ket{0},\ket{1}\}$ basis and classical control determined by the measurement outcomes. The Gottesman-Knill theorem tells us that the such stabilizer operations can be efficiently simulated classically (see for example \cite[Theorem~10.7]{NielsenChuang:2000}), while it also known that the addition of any other one-qubit gate outside the Clifford group will enable the system to perform \textsc{uqc}.  This fact provides us with a framework for testing tolerance to noise---one can examine how noisy this additional non-Clifford gate can be before it becomes classically simulable itself. If the non-Clifford operation has become a probabilistic combination of Clifford gates due to the noise, then we know that we are unequivocally in the classical computational regime. The noise rate where the extra gate becomes simulable (where it enters the ``Clifford polytope'' \cite{BuhrmanCleveLaurent:2006}) is an thus upper bound for fault tolerance. If the converse is true---i.e. if any operation outside the Clifford polytope enables \textsc{uqc} then the threshold is tight. In this article we show that for single-qubit gates undergoing depolarizing such a tight noise threshold does indeed apply. We will do so by proving that any depolarized gate that lies outside the Clifford polytope of single-qubit operations, in combination with noiseless stabilizing operations, allows for \textsc{uqc}.
This result should be contrasted to the situation for non-stabilizer qubit \emph{states} where the thresholds in that case are not currently known to be tight.

We will consistently assume that Clifford gates can be implemented perfectly, motivated by the fact that these gates can be implemented fault-tolerantly by applying them transversally and to encoded states \cite{CalderbankShor:1996,Steane:1996,Gottesman:1998,Knill:2004}. The fault-tolerant implementation of Clifford gates naturally carries with it a threshold of its own, independent of the kind we discuss in this paper.  The current model is particulary relevant to the so-called Pfaffian quantum Hall state in topological quantum computation \cite{FreedmanNayakWalker:2006,Georgiev:2006}, where the whole Clifford group (but only the Clifford group) can be implemented using braiding making these operations naturally fault-tolerant. The additional resource required to perform \textsc{uqc} will likely be highly noisy, and so we can see the parallels with our model.

We will begin by listing a couple of previously known results in this area. Next, we will discuss the connection between the geometry of the Clifford polytope and stabilizer measurements, and show that tightness of a magic-state distillation procedure for single-qubit states automatically ensures tight thresholds for non-Clifford gates undergoing any kind of unital noise. Finally we show that currently known magic-state distillation techniques are sufficient to prove tight thresholds for a non-Clifford gate undergoing depolarizing noise.

\begin{figure}
\begin{center}
\epsfig{file=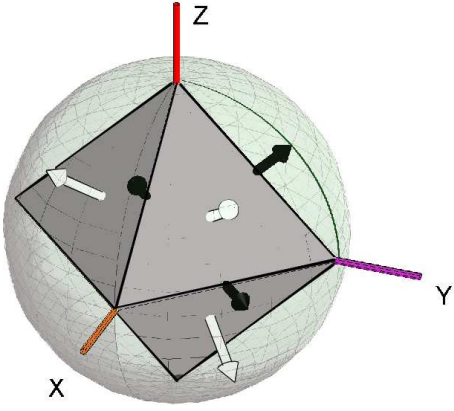,height=55mm}
\end{center}
\caption{\label{octo}%
 {Magic States and the Octahedron:} Some of the single-qubit
  magic states: $\ket{H}$ type states are designated with black
  arrows, $\ket{T}$ type states with white arrows. The octahedron
  defined by $|x|+|y|+|z|\leq 1$ depicts the single qubit states that
  can be created by stabilizer operations. Reichardt \cite{Reichardt:2005} showed that distillation techniques work right up to the edges of the octahedron (i.e. tight in the $\ket{H}$ direction). Current distillation techniques are unable to distill states just outside the faces of the octahedron (i.e.\ not tight in the $\ket{T}$ direction).}
\end{figure}

\paragraph*{Previously known results.} The idea of using perfect stabilizer operations in conjunction with imperfect non-stabilizer states to perform \textsc{uqc} originates with Bravyi and Kitaev \cite{BravyiKitaev:2005}. The conditions on the ancillary qubits to enable \textsc{uqc} is
that they are sufficiently close to being
one of the $20$ so-called magic states that lie on the surface of the
Bloch sphere.  The two classes of magic state (see Figure~\ref{octo}) are the $\ket{H}$ type and  $\ket{T}$ type, where all $\ket{H}$ type states can be derived by applying a Clifford operation to some canonical representative (and similarly for $\ket{T}$ states);
\begin{align}
\ket{H}&=\frac{1}{\sqrt{2}}\left(\ket{0}+\e^{\frac{\i\pi}{4}}\ket{1}\right)\\
\ket{T}&=\cos(\vartheta)\ket{0}+\e^{\frac{\i\pi}{4}}\sin(\vartheta)\ket{1},\quad\cos(2\vartheta)=\frac{1}{\sqrt{3}}
\end{align}
The routines used in \cite{BravyiKitaev:2005} were unable to distill qubit states just outside the edges and faces of the octahedron of Figure~\ref{octo}. Reichardt \cite{Reichardt:2005} subsequently suggested an improved routine that closed the gap in the $\ket{H}$ direction (along the edges of the octahedron).
Virmani et al.\ \cite{VirmaniHuelgaPlenio:2005} suggested using the convex hull of Clifford operations in order to find gates' robustness to various types of noise. In particular they considered gates that are diagonal in the computational basis. Plenio and Virmani \cite{PlenioVirmani:2008} subsequently extended this idea by analyzing cases where noise was allowed to affect the stabilizer operations too. Buhrman et al.\ \cite{BuhrmanCleveLaurent:2006} used a similar idea (that noise causes non-Clifford gates to eventually become implementable via Clifford gates only) to find the non-Clifford gate that is most resistant to depolarizing noise---a $\pi/8$ rotation about the $Z$ axis (or the same gate modulo some Clifford operation). Reichardt \cite{Reichardt:2006} showed that this particular gate enabled \textsc{uqc} right up to its threshold noise rate (about $45\%$), as well as considering in detail the process of reducing multi-qubit states to single-qubit states using postselected stabilizer operations. Our current result here generalizes this tightness result to all possible single-qubit gates.

\paragraph*{Preliminaries and Notation.} Let us parameterize an arbitrary single-qubit $\SU(2)$ gate as follows
\begin{equation}
U(\theta,\gamma,\delta)=\left(%
\begin{array}{cc}
  \e^{\i\gamma}\cos(\theta) & -\e^{\i\delta}\sin(\theta)  \\
  \e^{-\i\delta}\sin(\theta) &  \e^{-\i\gamma}\cos(\theta)\\
\end{array}%
\right)
\end{equation}
The representation of this rotation in $\SO(3)$ is denoted by $R(\theta,\gamma,\delta)$.
Implementing a rotation $R$ while suffering depolarizing noise (with noise rate $p$),
means that this noisy operation is represented by the rescaling $M=(1-p)R$,
a fact that we will need later.

Often we will apply the unitary $U(\theta,\gamma,\delta)$ to one half of an entangled Bell pair, $\ket{\Phi}=\frac{1}{\sqrt{2}}(\ket{00}+\ket{11})$, yielding
\begin{equation}
\varrho=(I\otimes U)\dm{\Phi}{\Phi}(I\otimes U)^\dag \label{vardef}
\end{equation}
If we use the two-qubit Pauli operators as a basis for the density matrix $\varrho$ then we can find the $16$ real coefficients $\text{c}_{i j}=\text{Tr}\left(\varrho (\sigma_i\otimes\sigma_j)\right)$ so that
\begin{equation}
\varrho=\frac{1}{4}\sum \text{c}_{i j}(\sigma_i\otimes\sigma_j) \qquad i,j \in \{I,X,Y,Z\}.
\end{equation}
Since we have applied a local unitary to a maximally entangled state, the coefficients $(\text{c}_{IX},\text{c}_{IY},\text{c}_{IZ},\text{c}_{XI},\text{c}_{YI},\text{c}_{ZI})$ are always zero. Comparing the $9$ coefficients $\left\{\text{c}_{XX},\text{c}_{XY}, \dots, \text{c}_{ZZ}\right\}$ one can see that these are the same as the entries of the $\SO(3)$ matrix $R(\theta,\gamma,\delta)$. More precisely,
\begin{equation}
R(\theta,\gamma,\delta) = \Bigg(
                    \begin{array}{rrr}
                      \text{c}_{XX} & -\text{c}_{YX} & \text{c}_{ZX} \\
                      \text{c}_{XY} & -\text{c}_{YY} & \text{c}_{ZY} \\
                      \text{c}_{XZ} & -\text{c}_{YZ} & \text{c}_{ZZ} \\
                    \end{array}
                  \Bigg)\label{correspondence}
\end{equation}
where the $c_{i j}$ are obviously also functions of $(\theta,\gamma,\delta)$.

If we represent the $24$ single-qubit Clifford operations as $\SO(3)$ matrices, then they are simply signed permutation matrices with unit determinant (they are a matrix representation of the elements of the chiral octahedral symmetry group or, equivalently, the symmetry group $\text{S}_4$). We label these operations $C_i$ and so the convex hull of the $C_i$ (the so-called Clifford polytope) is given by
\begin{equation}
\mathcal{P} =
\Bigg\{\sum_{i=1}^{24}p_i C_i\quad\Bigg|\quad\mbox{with $p_i\geq0$ and~}\sum_{i=1}^{24}p_i=1\Bigg\}.
\end{equation}
Geometrically, the Clifford polytope is a closed polyhedron in $\mathbb{R}^{9}$ that has $24$ vertices (each vertex representing one of the $C_i$). This polytope can also be defined by the bounding inequalities of its $120$ facets. The concise description of these facets used by Buhrman et al.\ \cite{BuhrmanCleveLaurent:2006} is given by the set
\begin{equation}
\mathcal{F}=\{C_iFC_j|i,j \in \{1,\dots,24\}, F \in \{A,A^T,B\}\}
\end{equation}
where
\begin{equation}
A=\Bigg(%
\begin{array}{rrr}
  1 & 0 & 0 \\
  1 & 0 & 0 \\
  1 & 0 & 0 \\
\end{array}%
\Bigg)
\quad\text{and}\quad B=\Bigg(
       \begin{array}{rrr}
         0 & 1 & 0 \\
         1 & 0 & -1 \\
         1 & 0 & 1 \\
       \end{array}
     \Bigg).\label{facets}
\end{equation}
At times we will have reason to refer to different subsets of the set of facets $\mathcal{F}$ so we use the obvious notation
\begin{equation}
\mathcal{F}=\mathcal{F}_A\cup\mathcal{F}_{A^T}\cup\mathcal{F}_B.
\end{equation}
It is useful to note that all the facets derived from $A$ comprise a single column with $\pm1$ entries and zeros elsewhere, and similarly for the row facets derived from $A^T$, hence
$|\mathcal{F}_A|=|\mathcal{F}_{A^T}|=3\cdot2^3=24$.
There are $|\mathcal{F}_B|= 72$ ``$B$-type'' facets, which can be constructed as follows: (i) Pick one position in the matrix e.g.\ $M_{i,j}$ and put $\pm1$ there ($9\times2=18$ choices), (ii) Fill the remaining entries not in row $i$ or column $j$ with $\pm1$ such that $\det(M)=-2$ ($4$ choices).

To determine whether or not an operation $M$ is inside the Clifford polytope $\mathcal{P}$ we take the elementwise inner product (or Frobenius inner product) between $M$ and the facets $F\in\mathcal{F}$ of the polytope
\begin{equation}
M \cdot F=\sum_{i,j=1}^{3}M_{i,j}F_{i,j}=\text{Tr}(M^TF).
\end{equation}
Using the above notation, a $3\times3$ matrix $M$ is inside the polytope $\mathcal{P}$
if and only if for all $F\in\mathcal{F}$ we have $M \cdot F\leq 1$.

\paragraph*{Interpreting the facets of the Clifford polytope.} Our proof will involve applying some non-Clifford gate to one half of a Bell Pair (as in Eq.~\ref{vardef}) and then postselecting on the outcomes of various stabilizer measurements. This has the effect of taking our two-qubit state $\varrho$ to a single qubit state $\rho'$ (times some stabilizer state that we do not care about), which we then distill using magic state distillation (see \cite{Reichardt:2006} for a more general discussion of these kinds of techniques). For example, performing a $YX$ measurement on $\varrho$ and postselecting on a ``$+1$'' outcome (i.e. projecting with $\Pi=\frac{1}{2}\left(II+YX\right)$) leads to a single-qubit state $\rho'$ with a Bloch vector given by
\begin{equation}
\vec{r}(\rho')=\Big(0,\four{XZ}{-}{ZY}{II}{+}{YX},-\four{XY}{+}{ZZ}{II}{+}{YX}\Big).
\end{equation}
The form of this vector means that it lies in the $YZ$ plane (see Figure~\ref{octo}), where we know that distillation techniques work right up to the edge $|y|+|z|=1$ of the octahedron. We can check if $\vec{r}$ is outside the octahedron by simply comparing the $L^1$ norm of $\vec{r}$ with 1. Rearranged, the condition $||\vec{r}||_1>1$ for being outside the octahedron is
\begin{equation}
|\text{c}_{XZ}-\text{c}_{ZY}|+|-\left(\text{c}_{XY}+\text{c}_{ZZ}\right)|\geq |\text{c}_{II}+\text{c}_{YX}| \label{absval}.
\end{equation}
Given the correspondence between the coefficients $\text{c}_{i j}$ and the elements of $R$ (see Eq.~\ref{correspondence}) we can rewrite the above condition (dropping the absolute value operators) as a facet inequality
\begin{equation}
R\cdot F>1\quad \text{where}\quad F=\Bigg(
       \begin{array}{rrr}
         0 & 1 & 0 \\
         -1 & 0 & -1 \\
         1 & 0 & -1 \\
       \end{array}
     \Bigg).
\end{equation}
This facet is a legitimate ``$B$-type'' facet and a little thought shows that, had we applied the single qubit Pauli operations (as $\SO(3)$ rotations) $X,Y$ or $Z$ to $\vec{r}(\rho')$ above, we would arrive at three other ``$B$-type'' facets
\begin{equation}
{\Bigg(
       \begin{array}{rrr}
         0 & 1 & 0 \\
         1 & 0 & 1 \\
         -1 & 0 & 1 \\
       \end{array}
     \Bigg)},~
\Bigg(
       \begin{array}{rrr}
         0 & 1 & 0 \\
         1 & 0 & -1 \\
         1 & 0 & 1 \\
       \end{array}
     \Bigg), \mbox{~or~}
\Bigg(
       \begin{array}{rrr}
         0 & 1 & 0 \\
         -1 & 0 & 1 \\
         -1 & 0 & -1 \\
       \end{array}
     \Bigg).
\end{equation}
respectively. Note that all four facet inequalities combined could be simplified to the form Eq.~\ref{absval} above. We omit the details, but it should be easy to see that all $72$ ``$B$-type'' facets correspond to postselecting on some (weight two) Pauli operator, and possibly performing a single qubit Pauli rotation on the resulting $\rho'$.

It is somewhat more straightforward to see the geometrical interpretation of the ``$A^{(T)}$-type'' facets. For example, the canonical $A$ given in Eq.~\ref{facets}, merely returns the sum of the elements of the Bloch vector $\vec{r}$, arising from a rotation applied to the $X$ ``$+1$'' eigenstate.
\begin{equation}
R\cdot A=\sum_{i=1}^{3}r_{i}\quad \text{where}\quad
\vec{r}=R\Bigg(
                                                     \begin{array}{c}
                                                       1 \\
                                                       0 \\
                                                       0 \\
                                                     \end{array}
                                                   \Bigg).
\end{equation}
 In general, an operation $M$ having an inner product greater than one with some ``$A$-type'' facet simply means that $M$, applied to some initial vector corresponding to a stabilizer state, brings that vector to a final position outside the octahedron.

 The preceding discussion shows us that if magic state distillation was possible everywhere outside the octahedron, then every unital operation outside the Clifford polytope would be distillable --- either straightforwardly or by using postselection, depending on which facet it violated. Using current (not tight) distillation routines however, we would be unable to deal with some operations violating an ``$A$-type'' facet by a fairly small amount. In the next section we show that, for depolarizing noise, any noisy rotation violating an ``$A$-type'' facet also violates a ``$B$-type'' facet. Since ``$B$-type'' facets correspond to $\ket{H}$ state distillation, the results we obtain are tight.

\paragraph*{Tight threshold for depolarizing noise.}
The claim we shall prove is that, anytime a matrix $M=(1-p)R$, representing a depolarized rotation, is outside some ``$A$-type'' facet then there exists a ``$B$-type'' facet
that $M$ also lies outside. In fact, we will prove the slightly stronger statement
that for all $R\in\SO(3)$
\begin{align}
\forall A\in\mathcal{F}_A\cup\mathcal{F}_{A^T}, \exists B\in\mathcal{F}_{B}
\text{~such that~} R\cdot (B-A)\geq 0.
\end{align}

To simplify the proof we will repeatedly make use of the symmetries of the problem (see Eq.~\ref{facets}). We will pick a canonical ``$A$-type'' facet and assume that this gives the largest inner product with $R$ of all the $F\in \mathcal{F}_A$. If there was a larger inner product with some $F\in \mathcal{F}_{A^T}$, then we could just relabel $R^T$ as $R$. We can assume that the facet with ones in the first column (the $A$ in Eq.~\ref{facets}) gives the biggest inner product since
\begin{equation}
\text{Tr}(R^T (C_iAC_j))=\text{Tr}(C_jR^T C_iA)=\text{Tr}((C_kR C_l)^TA).
\end{equation}

The proof will hinge on an entry of $R$, outside of the first column, being larger than the rest of the elements outside the first column. As such, let us define $12$ matrices closely related to $A$ and call them $A^\prime_i$:
\begin{align*}
 & A^\prime_1=\Bigg(%
\begin{array}{rrr}
  1 & -1 & 0 \\
  1 & 0 & 0 \\
  1 & 0 & 0 \\
\end{array}%
\Bigg)
 & A^\prime_2=\Bigg(%
\begin{array}{rrr}
  1 & 1 & 0 \\
  1 & 0 & 0 \\
  1 & 0 & 0 \\
\end{array}%
\Bigg)\\
 & \cdots
 & A^\prime_{12}=\Bigg(%
\begin{array}{rrr}
  1 & 0 & 0 \\
  1 & 0 & 0 \\
  1 & 0 & 1 \\
\end{array}%
\Bigg)
\end{align*}
such that the index $i$ largest inner product $R\cdot A^\prime_i$ tells us the sign and location of the largest magnitude element outside the first column. Once again, symmetry allows us to assume that $A^\prime_1$ yields the largest inner product because the rest of the $A^\prime_i$ can be derived from $A^\prime_1$ via Clifford rotations
\begin{equation}
\{A^\prime_i\}=\Bigg\{\Bigg(%
\begin{array}{rrr}
  0 & 0 & 1 \\
  1 & 0 & 0 \\
  0 & 1 & 0 \\
\end{array}%
\Bigg)^j A^\prime_1 \Bigg(%
\begin{array}{rrr}
  1 & 0 & 0 \\
  0 & 0 & 1 \\
  0 & -1 & 0 \\
\end{array}%
\Bigg)^k\Bigg| \begin{array}{c}
                                             j\in\{1,2,3\} \\
                                              k\in\{1,2,3,4\}
                                           \end{array}
\Bigg\}.
\end{equation}

For a matrix $R$ to be an $\SO(3)$ rotation there are constraints on the signs of the elements $R_{i,j}$ i.e.\ there are $8$ choices for the first column, $6$ choices for the second column and $2$ for the third. Given that $A$ is the maximum facet for $R$, we have fixed the signs positively in the first column, reducing the number of types of rotation to $6\times2=12$. Since $A^\prime_1$ gives the maximum inner product with $R$ of all  $A^\prime_i$ we have that $R_{1,2}<0$, which reduces the number of rotation types further to $3\times2=6$. It can be shown that $R_{1,2}$ having larger magnitude than the rest of the elements $R_{i,j} (i\in\{1,2,3\},j\in\{2,3\})$ restricts the type of rotation further to one the following four types
\begin{equation*}
 R\in\left\{ \mbox{\scriptsize $\left(
                  \begin{array}{rrr}
                    + & - & + \\
                    + & + & - \\
                    + & + & + \\
                  \end{array}
                \right),
                \left(
                  \begin{array}{rrr}
                    + & - & - \\
                    + & + & - \\
                    + & + & + \\
                  \end{array}
                \right),
                \left(
                  \begin{array}{rrr}
                    + & - & - \\
                    + & + & - \\
                    + & - & + \\
                  \end{array}
                \right),
                \left(
                  \begin{array}{rrr}
                    + & - & + \\
                    + & - & - \\
                    + & + & + \\
                  \end{array}
                \right)$}\right\}.
\end{equation*}
This should not be surprising if one considers that
\begin{equation}
R_{1,2}=-(R_{2,1}R_{3,3}-R_{2,3}R_{3,1})
\end{equation}
because of the structure of $\SO(3)$ matrices, and the sign patterns listed above ensure $|R_{1,2}|$ is as large as possible.

We claim that the $B\in\mathcal{F}_B$ of Eq.~\ref{facets} will suffice
to prove the desired inequality $R\cdot(B-A)\geq 0$, which reads in matrix form
\begin{equation}
                \Bigg(
                  \begin{array}{rrr}
                    + & - & \cdot \\
                    + & \cdot & - \\
                    + & \cdot & + \\
                  \end{array}
                \Bigg)\cdot\Bigg(
       \begin{array}{rrr}
         -1 & 1 & 0 \\
         0 & 0 & -1 \\
         0 & 0 & 1 \\
       \end{array}
     \Bigg)\geq0.\label{keyineq}
\end{equation}
Using the relevant entries of $R$ we define a pair of $2$-vectors $\vec{u}$ and $\vec{v}$  as
\begin{equation}
\vec{u}=(R_{1,1},R_{1,2}) \quad \vec{v}=(R_{2,3},R_{3,3})
\end{equation}
then we can rewrite the above inequality Eq.~\ref{keyineq} as
\begin{equation}
||\vec{v}||_1-||\vec{u}||_1\geq0.\label{eq:normineq}
\end{equation}

\begin{figure}[t]
\begin{center}
\epsfig{file=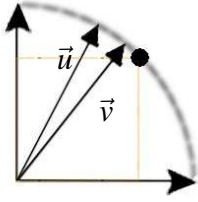,viewport=25 15 250 245 ,clip= ,height=30mm}
\rput[B]{1}(-2.2,1.7){ \large$\vec{u}$}
\rput[B]{1}(-1.65,1.0){ \large$\vec{v}$}
\end{center}
\caption{\label{proofpic}%
 {Proof of Eq.~\ref{eq:normineq}:} For any pair of 2-vectors $\vec{u}$ and $\vec{v}$ with the same $L^2$ norm, the vector with greater $L^\infty$ norm has smaller $L^1$ norm. A vector pointing towards the black dot has simultaneously minimal $L^\infty$ norm and maximal $L^1$ norm.}
\end{figure}

The $L^2$ normalization of all the rows and columns of the rotation matrix $R$ means that $\vec{u}$ and $\vec{v}$ have the same $L^2$ norm. With reference to Figure~\ref{proofpic}, it should be clear that because $\vec{u}$ has an $L^\infty$-norm at least as big as that of $\vec{v}$ (because $|R_{1,2}|\geq |R_{2,3}|,|R_{3,3}|$), it holds that the $L^1$-norm of  $\vec{v}$ is automatically at least as large as the $L^1$-norm of $\vec{u}$, as desired.

\paragraph*{Summary.}
We showed that for any unitary $1$-qubit gate undergoing depolarizing noise with rate $p$ it holds that if its $\SO(3)$ representation $M=(1-p)R$ lies outside the Clifford polytope $\mathcal{P}$, then it must be the case that there a facet $B\in\mathcal{F}_B$ such that
$M\cdot B > 1$. In turn, this means that if this noisy gate is applied to a Bell pair
$\ket{\Phi} = \frac{1}{\sqrt{2}}(\ket{00}+\ket{11})$ and an appropriate stabilizer measurement is applied, then, conditionally on the outcome of the measurement, one
obtains a state that can be transformed using Clifford gates into a single
qubit state with $|y|+|z|> 1$ in the Bloch ball representation. By the result of Reichardt \cite{Reichardt:2005} such states enable stabilizing operations to perform universal
quantum computation.

\paragraph*{Acknowledgments.}
This material is based upon work supported by the National Science Foundation.

\end{document}